\documentclass[sigplan,screen,authorversion,10pt]{acmart}
\usepackage{enumitem}
\usepackage{textcomp}
\usepackage{color,soul}
\usepackage{graphicx}
\usepackage{multirow}
\usepackage{pifont}
\usepackage{subfigure}
\newcommand{\cmark}{\ding{51}}%
\usepackage{amsmath,amsfonts}

\AtBeginDocument{%
  }

\usepackage{fancyhdr}
\AtBeginDocument{%
	\addtolength{\footskip}{2.0\baselineskip}%
	\fancyfoot[LE,LO]{\centering\begin{tabular}[b]{@{} c @{}}
			\textbf{\textit{Author's preprint version}} \\\strut
		\end{tabular}}
}
\copyrightyear{2023} 
\acmYear{2023} 
\setcopyright{acmlicensed}\acmConference[Middleware '23]{24th International Middleware Conference}{December 11--15, 2023}{Bologna, Italy}
\acmBooktitle{24th International Middleware Conference (Middleware '23), December 11--15, 2023, Bologna, Italy}
\acmPrice{15.00}
\acmDOI{10.1145/3590140.3592850}
\acmISBN{979-8-4007-0177-1/23/12}

\begin{document}

\title{Smartpick: Workload Prediction for Serverless-enabled Scalable Data Analytics Systems}

\author{Anshuman Das Mohapatra}
\orcid{0000-0002-5578-7109}
\affiliation{%
  \institution{University of Nebraska at Omaha}
  \streetaddress{1110 S 67th St}
  \city{Omaha}
  \state{Nebraska}
  \country{USA}
  \postcode{68182}
}
\email{adasmohapatra@unomaha.edu}

\author{Kwangsung Oh}
\orcid{0000-0003-3281-7325}
\affiliation{%
  \institution{University of Nebraska at Omaha}
  \streetaddress{1110 S 67th St}
  \city{Omaha, Nebraska}
  \country{USA}}
\email{kwangsungoh@unomaha.edu}

%
%
%
%
%

\renewcommand{\shortauthors}{Mohapatra et al.}

\begin{abstract}
	Many data analytic systems have adopted a newly emerging compute resource, serverless (SL), to handle data analytics 
	queries in a timely and cost-efficient manner, i.e., serverless data analytics.
	While these systems can start processing queries quickly thanks to the agility and scalability of SL, 
	they may encounter performance- and cost-bottlenecks based on workloads due to SL's worse 
	performance and more expensive cost than traditional compute resources, e.g., virtual machine (VM).
	In this paper, we introduce \textit{Smartpick}, a SL-enabled scalable data analytics 
	system that exploits SL and VM together to realize composite benefits, 
	i.e., agility from SL and better performance with reduced cost from VM.
	Smartpick uses a machine learning prediction scheme, decision-tree based Random Forest with Bayesian Optimizer, 
	to determine SL and VM configurations, i.e., how many SL and VM instances for queries, that meet cost-performance goals.
	Smartpick offers a \textit{knob} for applications to allow them to explore a richer cost-performance tradeoff space opened by exploiting 
	SL and VM together. 
	To maximize the benefits of SL, Smartpick supports a simple but strong mechanism, called \textit{relay-instances}. 
	Smartpick also supports event-driven prediction model retraining to deal with workload dynamics.
	A Smartpick prototype was implemented on Spark and deployed on live test-beds, Amazon AWS and Google Cloud Platform. 
	Evaluation results indicate 97.05\% and 83.49\% prediction accuracies respectively with up to 50\% cost reduction as opposed to the baselines. 
	The results also confirm that Smartpick allows data analytics applications to navigate the richer cost-performance tradeoff space efficiently and to handle workload dynamics effectively and automatically.
\end{abstract}




\maketitle

\section{Introduction}\label{sec:intro}
\subsection{Motivation}
Many Internet applications are running on cloud environments and generating large-scale data, 
e.g., Facebook \cite{facebook}, Twitter \cite{twitter} and Google \cite{google}. For these Internet applications, analyzing high 
volume of data is one of the most important workloads. For example, Facebook and Twitter analyze users’ posts, 
users’ activity logs, systems’ logs to query trends, make advertising decisions, and check overall cluster health.
Since the results of data analytics queries are usually used for making important decisions 
that affect revenues and system health, the queries must be processed promptly without a performance bottleneck.

To meet the performance goals, data analytics systems may deploy \textit{redundant} compute resources,
e.g., virtual machines (VMs), \textit{a prior}.
While this approach is simple and works well, this will incur additional cost (\$) for idle VMs. 
To avoid cost for unused compute resources, many previous 
works \cite{cherrypick, vanir, selecta, optimuscloud, optimus, ernest, paris, juggler, rpss, crystalLp} 
focused on determining optimal configurations, e.g., the number of VM instances and their types, and storage types, 
by predicting required compute resources for workloads. With these systems, additional VMs can be deployed to 
handle incoming queries without the performance bottleneck and idle VMs can be terminated to reduce cost based on workloads,
i.e., scalable data analytics systems. These systems, however, may not handle the \textit{latency-sensitive} 
queries promptly due to the unavoidable overhead of 
VM, i.e., boot-up latency ($>$ 55 seconds) \cite{coldBoot, coldBoot2}. 
If queries cause peak workload due to a lack of compute resources, 
they must wait until additional VM instances are fully deployed to be processed.

Many recent works \cite{spock, splitserve, occupy, flint, pocket, locus, numpywren, spark_on_lambda}
focused on adopting a newly emerging compute resource, \textit{serverless} (SL), such as Apache OpenWhisk \cite{openwhisk}, 
AWS Lambda \cite{lambdaRef}, Azure Functions \cite{azure_functions}, and Google Functions \cite {cloudFunctionsRef}, 
for data analytics to avoid the cold-boot latency problem, i.e., serverless data analytics (SDA).
Since SL offers \textit{agility}, very small boot-up time ($<$ 100 ms), and a \textit{pure-pay-as-you-go} cost model\footnote{Most popular cloud providers charge for SL only when the code is executed at either 1 millisecond (AWS) or 100 millisecond (GCP) granularity.}, SDA systems can deploy SL instances\footnote{We use the term serverless instances to refer serverless code invocations.} immediately and handle incoming 
queries without overprovisioned VMs in a cost-efficient way. 
These SDA systems, unfortunately, may still encounter cost- and performance bottlenecks based on data analytic workloads 
because SL offers worse performance and more expensive cost than VM \cite{cocoa,serverless_limitations,SLvsVMLee}.


\begin{table}
	\centering
	\caption{Comparison between SL and VM with the same amount of compute resources (2 vCPU with 2 GB RAM)}
	\scalebox{0.7} {
		\begin{tabular}{|c||c|c|}
			\hline
			& SL & VM \\
			\hline
			\hline
			Agility (Boot latency)& \textbf{High ($<$ 100 ms)}  & Low ($>$ 55 seconds) \\
			\hline
			Performance & Varying based on memory size &  \textbf{Relatively constant}\\
			\hline
			Cost Efficiency & \begin{tabular}{c}  \textbf{High (Pure pay-as-you-go :}\\\textbf{only when executed)}\end{tabular} & \begin{tabular}{c} Low (Pay-as-you-go \\: when deployed)\end{tabular}\\
			\hline
			Unit Time Cost (\$) & \begin{tabular}{c} Expensive (up to 5.8X) \end{tabular} &  \textbf{Cheaper}\\
			\hline
		\end{tabular}
	}
	\label{tab:compare}
\end{table}

Table \ref{tab:compare} shows the comparisons between SL and VM,
which represents different cost-performance points.
While data analytics systems may choose either one based on their resource demands and goals,
it would be highly desirable for them to achieve composite benefits (\textbf{bold} in Table \ref{tab:compare}), 
i.e., agility and cost-efficiency from SL and better performance and cheaper cost from VM.
However, determining compute resources configurations, e.g., how many SL and VM instances, 
is challenging due to the complexities: 1) heterogeneous compute resource characteristics,
2) workload prediction (how long a query will be executed), 3) diverse cost-performance goals,
and 4) dynamics from workloads.
While some recent works \cite{cocoa, splitserve, spock, libra, perf-cost-ratio, robustScaling} tried 
to exploit SL 
and VM together but they could not address these challenges as they have focused on 
either simple workload (independent tasks) or simple assumption without workload prediction.
Thus, they may not work well for data analytics.  

In this paper, we introduce \textit{Smartpick}, a serverless-enabled data analytics system (SEDA), 
that helps data analytics applications achieve desired cost-performance goals by addressing aforementioned challenges.
To determine \textit{optimal} cloud configurations of SL and VM, 
Smartpick uses a machine learning technique, decision-tree based Random Forest (RF) coupled with Bayesian Optimizer (BO), 
that predicts data analytic workloads using historical information.
Smartpick provides a \textit{knob} that allows applications to easily explore the cost-performance tradeoff space opened by exploiting SL and VM together.
Smartpick supports a simple but strong mechanism called \textit{relay-instances} to further improve performance with reduced cost.
To handle workload dynamics, Smartpick uses an event-driven approach that triggers a model retraining task to automatically evolve 
prediction models.

A Smartpick prototype implementation was built on the Spark \cite{sparkRef}, so that Spark applications can easily utilize
our system by setting diverse Smartpick's properties \textit{without any modification}.
We evaluated Smartpick on live-testbeds, Amazon AWS and Google Cloud Platform (GCP), using well-known benchmarks: TPC-DS \cite{tpcds}, Word Count \cite{wordCountInHive}, and TPC-H \cite{tpch}. 
Evaluations show that Smartpick can accurately characterize the TPC-DS workload performance with accuracies of 97.05\% on AWS and 83.49\% on GCP. 
The experimental results show that Smartpick can reduce cost by up to 50\% without performance degradation by using the relay-instances mechanism.
The results also confirm that Smartpick allows applications to easily explore the richer cost-performance tradeoff space with a simple knob and to handle workload dynamics by retraining the prediction model automatically. 


\begin{table}
	\centering
	\caption{Feature comparison with state-of-the-art. $\triangle$ indicates that metric is
		considered but with limitations}
	\scalebox{0.74} {
		\begin{tabular}{|c|c|c||c|}
			\hline
			& Cocoa & SplitServe & Smartpick \\
			\hline
			\hline
			Exploiting SL \& VM & \cmark  & \cmark & \cmark \\
			\hline
			Workload Prediction &  &  & \cmark \\
			\hline
			Handling Dynamics &  &  & \cmark \\
			\hline
			Segueing (Relay-instances) &   & $\triangle$ & \cmark \\
			\hline
			Cost-performance Tradeoff & $\triangle$ &  & \cmark \\
			\hline
		\end{tabular}
	}
	\label{tab:compareStateOfArt}
\end{table}

\subsection{Research Contributions}
Table \ref{tab:compareStateOfArt} compares Smartpick approach to two recent SEDA systems, i.e., Cocoa \cite{cocoa} and SplitServe \cite{splitserve}. 
While these systems utilize both SL and VM, they do not predict queries' workloads but just rely on external workload prediction systems \cite{cherrypick, vanir, selecta, optimuscloud, optimus, ernest, paris, juggler, rpss, crystalLp}.
However, these prediction systems may not work well in SEDA due to their SL-agnostic approach and workload dynamics, which significantly affect overall cost and performance.
Thus, we designed the workload prediction module to easily work with any SEDA system that needs performance prediction.
Since using SL for a long time would incur additional cost without performance improvement \cite{cocoa, splitserve},
Smartpick judiciously and dynamically terminates SL instances using the mechanism called \textit{relay-instances}. 
While SplitServe \cite{splitserve} uses a similar technique called segueing,
they use a static approach, which leads to significant cost inflation.
While Cocoa considers exploring the cost-performance tradeoff space like Smartpick, 
its performance is highly dependent on several static parameters that may be hard to tune in SEDA.

To summarize, the research contributions are as follows:
\begin{itemize}[wide = 0pt, topsep=0pt]
	\renewcommand{\labelitemi}{\noindent$\bullet$}
	\item The design and implementation of Smartpick, the first scalable data analytics system 
	(to the best of our knowledge) that predicts data analytics workloads with consideration of 
	SL and VM together to determine optimal compute resource configurations. 
	\item Flexibility that allows \textit{unmodified data analytics applications and other SEDA systems} to reap the benefits.
	\item A simple way to easily explore the cost-performance tradeoff space using diverse mechanisms embedded within the workload prediction.
	\item Event-driven re-training of the prediction model to handle workload dynamics, e.g., varying data size and new queries.
	\item Thoughtful empirical evaluations on AWS \cite{awsRef} and GCP \cite{gcpRef}, showing the efficacy of Smartpick. 
\end{itemize}

\section{System Model and Motivation}\label{back}
\subsection{System Model}\label{subsec:sysmodel}
\noindent\textbf{Data center (DC) setting and compute resources:} We focus on a single DC environment in the public cloud, where the network is not a performance bottleneck \cite{irre}
and infinite compute resources, i.e., serverless (SL) and virtual machine (VM), are available.
Each compute resource has different characteristics in terms of performance, cost, and agility, as shown in Table \ref{tab:compare}.
Such compute resources heterogeneity opens a rich cost-performance tradeoff space that applications can explore based on their demands.
While data within a DC can be accessed and processed without a performance bottleneck, achieving memory-locality is important for performance improvement \cite{irre}. 
Exploiting SL in data analytics requires external storage systems, e.g., Redis \cite{redis} or AWS S3 \cite{s3Ref}, due to its 
limitations, e.g., limited network and storage, which may incur performance overhead.
We assume that performance overhead from losing memory-locality is negligible as we target queries with several tens of seconds granularity. 
We will discuss potential performance improvement with improved memory locality in Section  \ref{sec:related}.

\noindent\textbf{Data analytics applications:} 
We consider data analytics applications that generate diverse classes of MapReduce-like queries, 
e.g., reporting, ad-hoc, iterative, and data mining, as classified in \cite{whytpcds, whytpcds2}.
These queries contain several map and reduce stages that  cannot start until 
all their dependencies are resolved, i.e., dependent tasks. 
These queries can be processed by de-facto distributed data processing frameworks, e.g., Hadoop \cite{hadoop} and Spark \cite{sparkRef}.
While reporting queries are somewhat predictable as they are regularly generated based on the schedule, i.e.,
recurring (\textit{static}) queries, the remaining classes of queries, especially ad-hoc queries, are impromptu and dynamically
constructed to answer immediate and specific questions, i.e., \textit{dynamic} queries.
In this work, we mainly consider dynamic queries that may cause peak workloads.
Applications may utilize infinite compute resources, e.g., redundant VM instances, 
to handle dynamic queries without the performance bottleneck, which incurs additional cost for under-utilized or idle compute resources \cite{underutil}.
We assume that they have limited operational budgets; thus, minimizing the cost of processing queries within their target performance goals is highly desirable.

\noindent\textbf{Data analytics system (DAS):}
We assume that DAS deploys an optimal number of long-lived VM instances as \textit{static compute resources} 
to handle static queries using workload prediction tools or systems \cite{cherrypick, vanir, selecta, optimuscloud, optimus, ernest, paris, juggler, rpss, crystalLp}.
However, DAS may encounter a performance bottleneck due to peak workloads (lack of compute resources) 
caused by the dynamic queries, e.g., ad-hoc queries.
While DAS can deploy additional VM instances to handle the dynamic queries, applications may not achieve the desired performance goals due to unavoidable overhead of VM, i.e., cold boot-up latency ($>$ 55 seconds)  \cite{coldBoot, coldBoot2}. 
Instead, DAS may deploy SL instances to start processing queries immediately as done in previous works \cite{spock, splitserve, occupy, flint, pocket, locus, numpywren, spark_on_lambda}, 
i.e., serverless data analytics (SDA). 
However, based on query workloads, SDA may encounter the cost-bottleneck for little (or no) performance improvement \cite{cocoa}. 
To handle dynamic queries in a timely and cost-efficient way,
we consider DAS that uses a \textit{hybrid approach exploiting SL and VM together} 
to achieve composite benefits, i.e., agility and cost-efficiency from SL, and better performance and cheaper cost from VM.

\noindent\textbf{Determining optimal compute resource configuration problem:}
While recent works \cite{cocoa, splitserve, spock, libra, perf-cost-ratio, robustScaling} have introduced 
similar hybrid approaches, they adhere to simple assumptions or 
workloads, e.g., static parameters without workload prediction, 
dynamics-free prediction model, and independent tasks, 
which would not work well for serverless-enabled data analytics (SEDA). 
In this work, we focus on determining the optimal compute resource configurations, i.e., 
how many SL and VM instances need to dynamically be spawned to handle incoming queries.
However, this is challenging because many metrics must be considered, 
e.g., query workload estimations (prediction), diverse applications' cost-performance goals, 
and heterogeneous compute resource characteristics. 
To determine optimal configurations, diverse approaches have been introduced to 
build performance prediction models using historical data \cite{cherrypick, vanir, selecta, optimuscloud, optimus, ernest, paris, juggler, rpss, crystalLp}.
Unfortunately, these systems do not consider SL, but only VM for compute resources and
thus do not work well for SEDA. 
Furthermore, with a large search space for optimality, novel approaches are required to navigate the solution space efficiently and ensure acceptable overhead/cost for the decision-making.
In this work, we use a machine learning technique, decision-tree based Random Forest (RF), to predict data analytic workloads using historical information. 
To efficiently explore the large search space, we incorporate Bayesian Optimizer (BO) into our prediction model, i.e., RF + BO (Section \ref{sec:workload_prediction}).
Given predicted workloads, we focus on minimizing cost while meeting target performance goals, i.e.,
exploring a cost-performance tradeoff space (Section \ref{subsec:tradeoff}). 

\noindent\textbf{Dynamics:} 
We assume that applications may send new queries unknown to DAS at any time. In addition, data size can be changed as more data is aggregated. To predict workload correctly, the prediction model must be updated by incorporating these changes (Section \ref{subsec:dynamics}).


\begin{figure*}[t!]
	\subfigure[100 tasks (SL-only is best)]{\label{subfig:illex1}\includegraphics[width=2in]{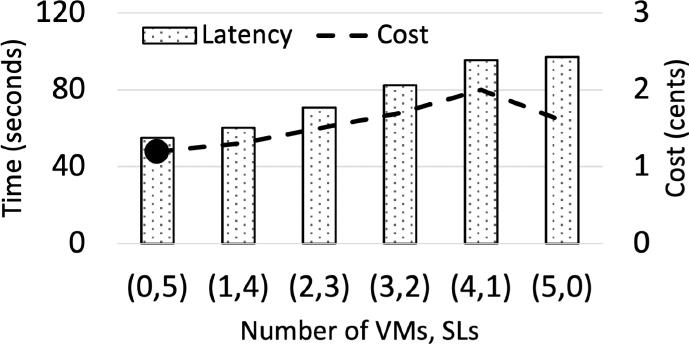}}
	\hfill
	\subfigure[250 tasks (Hybrid is best)]{\label{subfig:illex2}\includegraphics[width=2in]{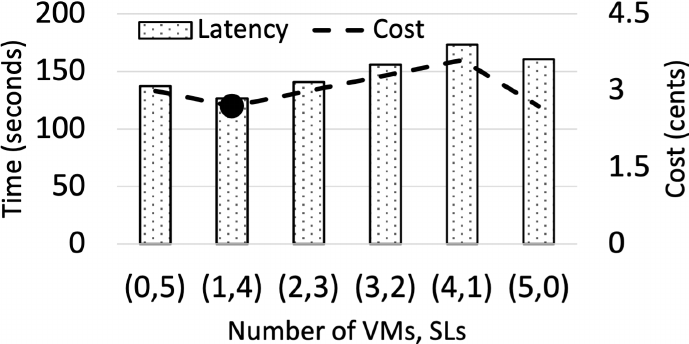}}
	\hfill
	\subfigure[500 tasks (Hybrid is best)]{\label{subfig:illex3}\includegraphics[width=2in]{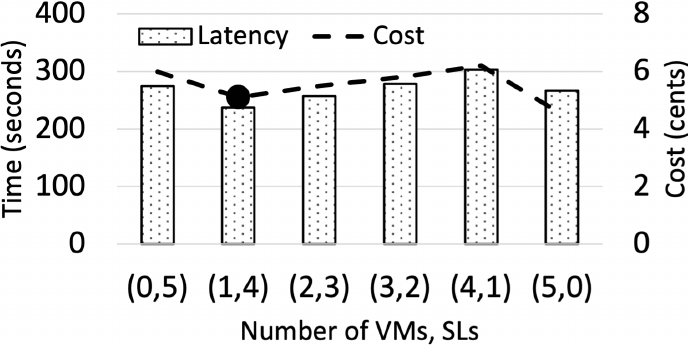}}
	\caption{Exploring resource determination and tradeoff. $\bullet$ indicates best performance.}
	\label{fig:allConfEx}
	\Description{This figure highlights the need for workload prediction by considering several configurations accross 3 different task sizes: 100 (short), 250 (mid) and 500 (long). SL-only is the preferred approach for short running queries, where as hybrid approach yields better performance for mid and long running queries. Moreover, this figure illustrates the existence of rich cost-performance tradeoff space, that can be explored for meeting finer application needs based on the budgeted requirements.}
\end{figure*}

\subsection{Illustrative Example} \label{subsec:example}
Workloads in data analytics systems (DAS) have large variance on query completion times. This stems from the fact that each of them can have different query semantics and thus, dissimilar resource needs to process the given data. To account for such scenarios and to handle the incoming queries efficiently, we highlight the need for performance prediction through an interpretative example. Let's assume three classes of dynamic queries: short-, mid-, and long-running queries, that incur peak workload. 
These queries have 100 tasks (short), 250 tasks (mid), and 500 tasks (long) respectively. Since all static compute instances are busy handling regular queries, DAS needs to deploy additional compute resources to handle them. In this case, DAS must determine optimal configurations, i.e., \textit{how many SL and VM instances}, 
that meet the applications' cost-performance tradeoff preference.

DAS has three options to deploy compute resources: 1) SL instances only (SL-only), 
2) VM instances only (VM-only), and 3) both VMs and SLs (Hybrid). For the sake of comparison, we consider AWS t3.small instance (2 vCPUs and 2 GB memory)
and AWS Lambda with 2GB memory. Note that AWS Lambda (2 GB) offers 2 vCPUs for each invocation. We take cost information from AWS \cite{t3Price, lambdaPrice}. 
We consider storage cost for each VM (gp2 8 GB) and Redis \cite{redis} (external) storage cost (on master VM instance) whenever SL instances are involved.
Note that we choose AWS t3 family for the same compute resources as SL instance, and we consider the burstable costs (\$0.05 per vCPU-hour) in our model. For the performance of SL instances, we assume zero-boot latency and include 30\% 
performance overhead to task execution time (based on experimental evidence as shown in Section \ref{subsec:setup}). For the VM-only approach, we added 55 seconds to the query completion time as the cold-boot overhead \cite{coldBoot, coldBoot2}.

Figure \ref{fig:allConfEx} presents the expected query execution time and cost when 
DAS applies different approaches for an incoming query, assuming that 5 instances (either SL, VM or combined) are the optimal number of CPU cores. Here (0,5) and (5,0) represent the two extremes of compute resources configuration, i.e., SL-only and VM-only approach, respectively.
For the short-query, the SL-only approach offers the best performance with reduced cost, thanks to the agility of SL. 
For mid- and long-queries, however, the SL-only approach inflates cost without performance improvement, while the hybrid approach leads to better performance with the average cost.
Interestingly, the VM-only approach outperforms the SL-only approach for long-running query due to the heterogeneity between SL and VM, as discussed in Section \ref{sec:intro}. 
The results clearly show that a workload prediction scheme is extremely important to determine the optimal configurations of VMs and SLs for varying query classes.
The results also indicate that there is a richer cost-performance tradeoff space based on the query workloads. 

\noindent\textbf{Relaying workload:} 
In the hybrid approach, SLs can be invoked and used until a query is completed, which may incur additional cost without performance improvement due to SL's characteristics, as discussed in Section \ref{subsec:sysmodel}.
To avoid this, SLs can be terminated when corresponding VM instances are ready to avoid cost inflation and performance degradation, i.e., \textit{relay-instances} mechanism. 
For example, for a long-running query (500 tasks), 5 SLs and 5 VMs can be allocated simultaneously. 
The 5 SLs start running the tasks quickly and will be terminated when the corresponding 5 VMs are ready for the rest of the tasks, i.e., after VM's cold-boot time.
This approach results in performance improvement to 198.8 seconds with a reduced cost of 5\textcent, which is a better approach than simply using SLs throughout the query execution. 
We will discuss the relay-instances mechanism in Section \ref {subsec:relay} in detail.
%
%
%
%
%
%
%
%
%
\section{Determining Optimal Configurations}\label{sec:workload_prediction}
\subsection{Workload Prediction} \label{subsec:wrkPredDetails}
While many workload prediction systems have been proposed \cite{ernest, vanir, selecta, paris, optimus, cherrypick, optimuscloud, crystalLp, rpss, juggler}, 
none of these works have considered SL to determine compute resource configurations. 
In this section, we introduce how Smartpick predicts query workload to determine the optimal configuration.

\noindent\textbf{Feature Determination}:
Precisely predicting the query completion time is one of the key aspects of Smartpick. To this end, we thoroughly analyzed what parameters uniquely determine query completion time. Based on multiple initial runs, we deduced the rich set of features that govern this behavior, which are summarized in Table \ref{tab:fea-wp}. When new queries are submitted to an already trained model, the \textit{query-duration} feature will act as the best estimation for completion time. Likewise, different \textit{instances} will be traversed, and the best combination of VMs and SLs will be determined for efficiently executing a new incoming job. 
Having determined the features, we next explored several approaches \cite{ernest, optimus, fim} for modeling these parameters into query completion time, however, all of these approaches rely heavily on the implicit relationship across the parameters, which can be very difficult to model. Therefore, in our design, we incorporate black-box model for optimal compute-resource determination.

\noindent\textbf{Problem Formulation}:
We choose decision-tree based Random Forest (RF) technique for quantifying the query completion time. This is preferred over other deep learning neural networks because it is computationally less intensive and requires significantly less training data \cite{rf-inexp, rf-inexp2, rf-inexp3, rf-inexp4}. Moreover, it reduces model over-fitting through the technique of ensemble learning \cite{rf-red-overfit}. Equation \ref{eq:bl} provides the formulation for the RF regressor, where $\beta$ is the rich set of identified features and \textit{$RF_t$} is the expected completion time.  
\begin{equation}
	f(\beta)=RF_t
	\label{eq:bl}
\end{equation} 
Although this regressor can accurately model the underlying system, the search space involved for exhaustive navigation is huge. Our initial experiments show around \textit{1 minute} of prediction latency when both VMs and SLs are involved for optimality determination. Given the time-sensitivity of data analytics workloads, exhaustive search proves a hindrance for efficient model performance. Therefore, we add a Bayesian Optimizer (BO) module to navigate the search space effectively. The BO in its raw form cannot be used for workload prediction of ad-hoc queries since this leads to a significant compute cost for the resource determination. We discuss these challenges in detail in Section \ref{RFBOAdv}. Hence, we modify the BO technique to tune it in accordance with cost-effectiveness. 

Two primary components are associated with the BO, i.e., objective and surrogate functions. Equation \ref{eq:ob} defines the objective function which is tailor-made for Smartpick. In this equation, $RF_t$ is the predicted query completion time from the RF regressor and $\delta$ is the noise value which follows normal distribution. The surrogate function is chosen to be a \textit{Gaussian Process Regressor}, since they demonstrate several remarkable characteristics. First, the variance in prediction accurately models the noise in observations, and second, it can precisely generate values for newer data points \cite{gps}.
\begin{equation}
	maximize: - (RF_t + \delta)
	\label{eq:ob}
\end{equation} 
For the acquisition function, there are several choices - Expected Improvement (EI), Probability of Improvement (PI) and Upper Confidence Bound (UCB) \cite{acq-types}. For Smartpick, we incorporate PI over the other options because it is similar to EI and simpler \cite{pipref}, as well as, it is one of the most widely used acquisition functions for optimizers \cite{widelyused-acqs}. Thus, PI helps in efficiently exploiting/exploring the search space for optimal/near-optimal compute resource configurations in the form of tuples: $\{nVM, nSL\}$, where nVM is the desired number of VMs and nSL is the 
desired number of SLs. The termination criteria of the search are aligned with the improvement to (estimated) query completion time. If the improvement does not increase by 1\% for 10 consecutive searches, the model returns the accomplished core configurations for VMs and SLs.

\begin{table}
	\centering
	\caption{Features for Workload Prediction}
	\label{tab:fea-wp}
	\scalebox{0.7} {\begin{tabular}{ll}
			\hline
			Feature&Comments\\
			\hline
			\textbf{instances}& Number of VMs and SLs used\\
			input-size& Size of input in bytes\\
			start-time-epoch& Initial job submit time in epoch\\
			total-memory& Total memory of available workers\\
			available-memory& Available memory of available workers\\
			memory-per-executor& Memory assigned to each executor\\
			num-waiting-apps& Number of applications in wait state\\
			total-available-cores& Number of available cores\\
			\textbf{query-duration} & Completion time of a given query\\
			\hline
	\end{tabular}}
\end{table}

\begin{figure}[!t]
	\centering
	\includegraphics[width=1.61in]{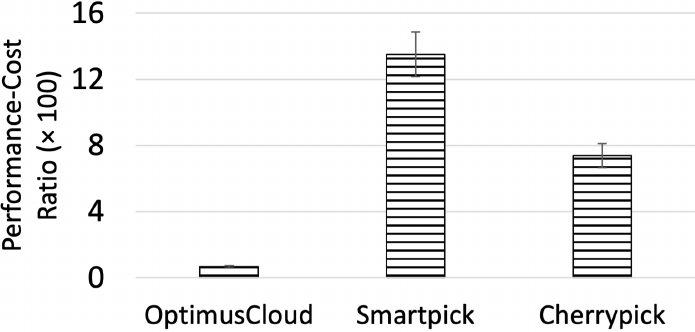}
	\caption{Comparison with known resource determination techniques (higher is better)} 
	\label{fig:allComp}
	\Description{This figure compares the proposed prototype with 2 other related works - OptimusCloud and Cherrypick. The primary vertical axis represents performance-cost ratio and overall observation suggests that Smartpick outperforms the other 2 state-of-the-art techniques.}
\end{figure}

\subsection{Why RF + BO is better than others?}\label{RFBOAdv}
Techniques proposed in latent factor collaborative filtering \cite{selecta}, 
machine learning models \cite{paris}, online fitting \cite{optimus},
Bayesian optimization \cite{cherrypick}, sampling \cite{ernest}, 
and a mix of other tools \cite{vanir} - work great when the search space involves only one type of compute resource (i.e., VMs). Some recent works utilized RF and BO to predict the workloads, 
e.g., OptimusCloud \cite{optimuscloud} uses RF and CherryPick \cite{cherrypick} uses BO. 
Since they considered a single instance type as compute resource, they may
simply add SLs as a new instance type in order to incorporate them. 
This approach, however, will lead to a huge search space for optimality, 
which \textbf{cannot} be traversed in a timely and cost-efficient way 
as they use RF and BO separately. 
To understand the benefits of the RF + BO approach, we tune 
our prediction model for OptimusCloud (RF-only) and CherryPick (BO-only) 
to incorporate both VMs and SLs. 
To compare different approaches, i.e., RF-only, BO-only, and RF + BO, 
we use performance-cost ratio ($PC_r$) \cite{perf-cost-ratio} that can be 
computed as shown in Equation \ref{eq:perfCost}. Here, \textit{Time} denotes the inference latency, whereas \textit{cost} denotes the compute charges incurred for model creation.
\begin{equation}
	PC_r = \frac{1/Time}{1+cost}
	\label{eq:perfCost}
\end{equation}
We put same inputs (features) to each prediction model 10 times to see
how each model works. 
Figure \ref{fig:allComp} shows our preliminary simulation results that is scaled to a multiple of 100 (higher
is better). It is evident that OptimusCloud \cite{optimuscloud} gives the worst $PC_r$ value because of the large overhead arising from search complexity. Moreover, CherryPick \cite{cherrypick} has better search complexity because of the surrogate design (of BO) but incurs a higher cost from the projected execution runs on live VM and SL instances. Overall, we observed the best $PC_r$ values for Smartpick since it not only reduces the search time complexity but also incurs a lower cost from the enhanced \textit{RF + BO} approach.

\subsection{Optimal Configurations with Preferences}\label{subsec:tradeoff}
Although optimal resource determination leads to minimum query latency, this may not be feasible for some applications that are sensitive to budget requirements. For these applications, some additional query latency 
would be tolerable for reducing operational cost, i.e., cost-performance tradeoff. 
Therefore, Smartpick supports a cost-performance tradeoff knob ($\epsilon$) that can be tuned as per the application's target cost-performance goals. 
Given the knob, Smartpick may proportionally scale down the determined SLs and VMs.
For example, setting the $\epsilon$ value to 0.5 halves the numbers of SL and VM instances from the optimal configurations determined for best performance.
While this approach is simple, we observed that this would lead to significantly high query completion times without a smoother navigation of cost-performance tradeoff.

Instead, Smartpick optimizes resource determination based on the tolerance level set i.e., $\epsilon$. 
Smartpick uses a list of estimated times ($ET_l$) to track the candidate solutions explored for the final optimum. This list is traversed before the final resource determination to meet desired cost-performance goals. Equation \ref{eq:troff2} shows the objective function that is modeled for finer and more precise control of tradeoff; $T_{est.}$ is the estimated time under consideration, $t_{vm}$ is the estimated VM time, $t_{sl}$ is the estimated SL time, $C_{vm}$ denotes compute cost per instance of VM, $C_{sl}$ denotes compute cost per instance of SL, $C_{best}$ is the cost value associated with optimal configuration and $T_{best}$ is the optimum time determined by Smartpick. 
\begin{equation}
	\begin{aligned}
		\max_{t} \quad & T_{est.}; \hspace{0.5em} T_{est.} \hspace{0.5em} \in \hspace{0.5em} ET_l\\
		\textrm{s.t.} \quad & {nVM} \times t_{vm} \times C_{vm} + {nSL} \times t_{sl} \times C_{sl} \leq C_{best}\\
		&T_{best} \times (\epsilon + 1) \geq T_{est.}\\
	\end{aligned}
	\label{eq:troff2}
\end{equation}

It aims to find higher query estimation times ($T_{est.}$) that is within the specified limits, i.e., tolerable additional latency ($2^{nd}$ constraint), 
but draws minimum compute cost ($1^{st}$ constraint). For instance, $\epsilon$ = 0.2 specifies a tolerance level of 20\% above the optimum value ($T_{best}$), but the actual cost could be lower for a reduced query latency. 
This is not always guaranteed though and the optimization problem helps ascertain the required values as shown in Section \ref{subsec:tradeoffspaceEval}.



\section{Smartpick Overview} \label{sec:design}
In this section, we present an overview of Smartpick.


\subsection{Smartpick Architecture \& Workflow}\label{subsec:workflowArch}
\begin{figure}[t!]
	\centering
	\includegraphics[width=2.7in]{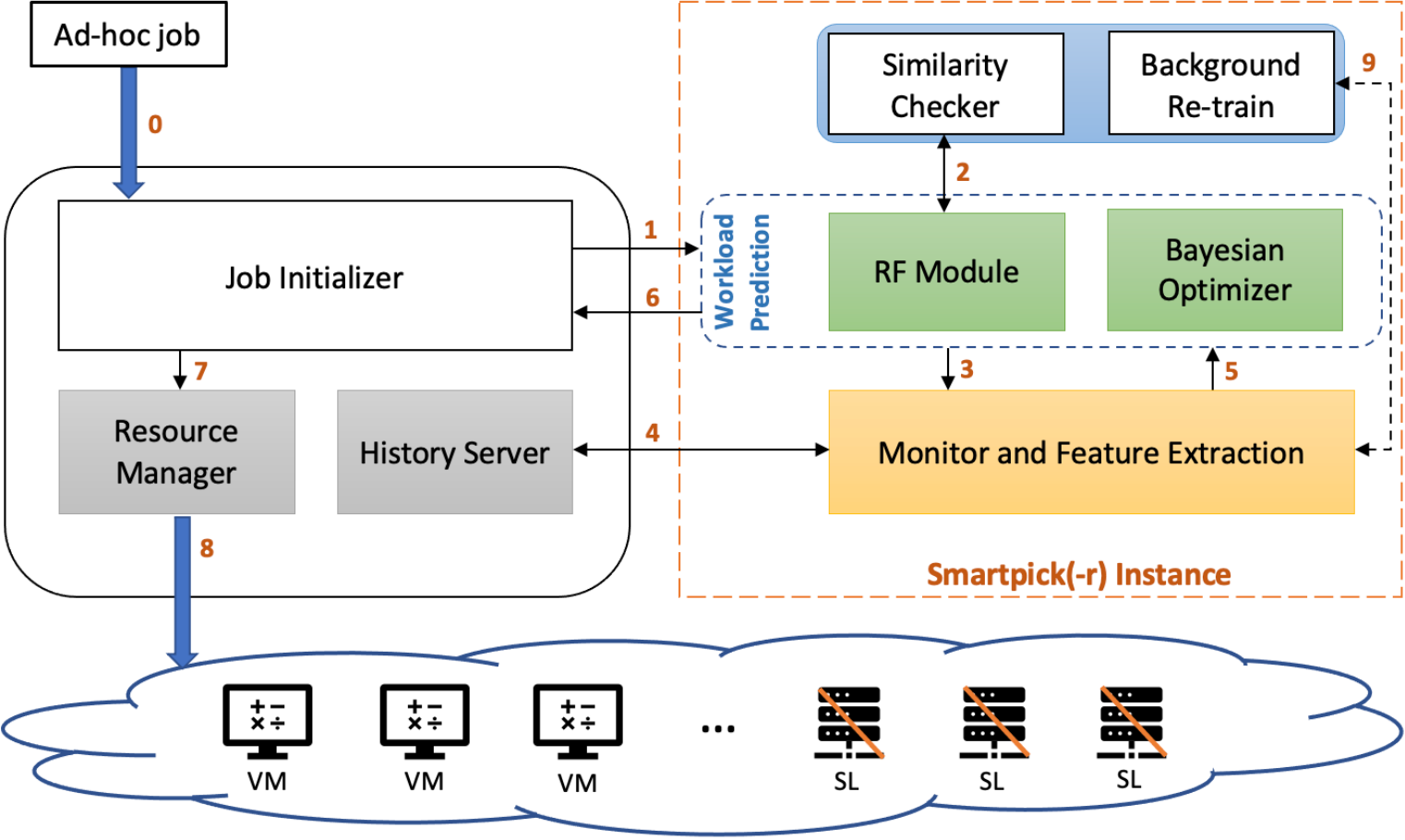}
	\caption{Smartpick Architecture}\label{fig:arch}
	\Description{This figure presents the overview of Smartpick architecture. It consists of two primary components – Workload Prediction and Elastic Module. These components along with the Monitor and Feature Extraction module work towards determination of optimal compute resources for an ad-hoc job. Based on the decisions provided by Smartpick, the Resource Manager spawns the required number of VMs and SLs in the cloud and the system finally records the event logs in History Server for background re-training tasks.}
\end{figure}

Figure \ref{fig:arch} shows the Smartpick architecture in which the numerical values show the order of query execution when a new query is sent to Smartpick.

\begin{itemize}[wide = 0pt, topsep=0pt]
	\renewcommand{\labelitemi}{\noindent$\bullet$}
	\item The workload prediction (WP) component comprises two sub-modules, i.e., RF and BO, that work together to efficiently explore the large search space as discussed in Section \ref{sec:workload_prediction}.
	\item Similarity Checker (SC) parses the alien (unknown) queries for extracting meaningful information such as the number of tables, columns, and subqueries inferred in the request.
	\item Monitor and Feature Extraction (MFE) monitors job execution, and 
	maintains a trained RF model and query features.
	\item History Server (HS) captures and stores the metrics outlined in Table \ref{tab:fea-wp}. 
	\item Resource Manager (RM) spawns and manages SL and VM instances based on optimal compute resource configurations. 
	\item Background Re-train creates a new model when the current model is outdated due to workload dynamics. 
\end{itemize}	

\noindent\textbf{Workflow:} When a new query is received (step 0), \textit{Job Initializer} (JI) asks WP to determine the optimal number of VMs and SLs required for the job (step 1). 
To efficiently predict query workload, WP maintains a list of queries against which the current operating model is trained. 
If WP realizes that the incoming query is not in the queries list, i.e., unknown query, WP asks the SC
to find the closest query/workload (in step 2). To determine optimal configurations, WP needs a trained RF model and 
query features as inputs except for \textit{instances} and \textit{query-duration}, as explained in Section \ref{subsec:wrkPredDetails}.
WP acquires these inputs from MFE (step 3) that pulls historical data from the History Server (step 4).
When all the inputs are available (step 5), WP can determine the optimal number of SLs and VMs. If the cost-performance tradeoff knob ($\epsilon$)
is set to greater than 0, WP iterates the Estimated Time list (or $ET_l$) to find a configuration that meets 
the cost-performance goal as explained in Section \ref{subsec:tradeoff}. 
From our evaluation, WP can determine compute configuration asynchronously (without blocking the Spark \cite{sparkRef} execution flow) within 
1.5 seconds for a known query and less than 2.5 seconds 
for an unknown (alien) query. We assume that this overhead is ignorable 
as we consider queries that take several tens of seconds. 
WP returns the resource requirements of incoming query to JI (in step 6). 
JI asks RM to spawn VMs and SLs based on the determination (step 7). 
RM spawns the desired number of VMs/SLs on the chosen cloud provider (step 8), following which the query execution begins.
If the prediction error in query execution (examined by MFE on job completion in step 9) is higher than the threshold, the prediction model is retrained by Background Re-train. 

\subsection{Handling Dynamics}\label{subsec:designDynamics}
\label{subsec:dynamics}
Workload dynamics could occur due to several reasons. 
For example, data analytics applications may need to write new queries to meet their needs \cite{changingNeeds}.
In addition, applications on the cloud store data in enormous volumes for decision-making and health checks \cite{bigDataForAnalysis},
i.e., increased data size. Smartpick is designed to handle such dynamics automatically.


\noindent\textbf{Similarity check for alien queries:} Determining compute resources for alien queries is challenging since the prediction model is completely unaware of their resource needs.
To make a reasonably accurate prediction for such unknown queries, 
Smartpick maintains the known queries' identifiers and their attributes, 
such as the number of tables, columns, subqueries, and map tasks. When queries are sent, Smartpick extracts these attributes from the incoming queries 
and computes the \textit{spatial cosine similarity} to search for the closest known-query identifier. This reference identifier, along with other inputs (as discussed in Section \ref{subsec:wrkPredDetails}), is then used to deduce the request's resource-needs. We will show that Smartpick with similarity 
can help achieve good performance with reduced cost for similar yet alien queries in Section \ref{subsubsec:simCheckEvalChild}.

\noindent\textbf{Retraining prediction models:} 
While Similarity Checker works well for alike queries, it does not account for workloads that are completely different from the trained queries. Thus, in the event of new/changed workloads, that is, when the accuracy is below an acceptable threshold, we need to retrain the prediction model. To achieve this, Smartpick monitors the difference between actual- and predicted- query execution time. If the difference is greater than a specified threshold, then Smartpick will spawn an asynchronous model re-training task that will re-tune the prediction models (in background) for handling dynamics.
In addition, this re-training needs to be highly configurable so that any application with specific needs can reap the maximum benefits out of it. 
We will discuss these configurable options in detail in Section \ref{sec:impl}. 
\subsection{Relay Instances} \label{subsec:relay}
To reap the benefits from the hybrid approach, i.e., SL + VM, they should be used in coordination.
This is because utilizing SL instances until when a query is completed may incur an additional cost with little 
(or no) performance improvement due to SL's more expensive cost and worse performance than VM, as discussed in Section \ref{sec:intro}.
To avoid this, Smartpick uses a simple but efficient mechanism, \textit{relay-instances}, with which 
the SL instances start running the tasks quickly, and will be terminated when corresponding VMs are ready for the rest of the tasks.
That is, SLs are only used during the VM's cold-boot time, and then terminated 
to maximize the benefits of the hybrid approach, i.e., agility from SL and better performance with reduced cost from VM. Consequently, Smartpick's prediction model incorporates the relay-instances mechanism, and thus, the VM and SL resources determined (which may be unequal but optimal) account for these relaying workloads.

SplitServe \cite{splitserve} offers a similar approach, called segueing. However, their approach 
relies on a static threshold to terminate SLs, which may be costly with limited performance improvement. 
In addition, they use the same numbers SL and VM, which may not be optimal for a query. 
For example, SLs can be idle during the static timeout in segueing, which inflates overall cost significantly with limited performance improvement.
We present the benefits of relay instances and cost-performance comparison between relay instances and segueing in Section \ref{subsec:eval_compare}.

\section{Smartpick Implementation} 
\label{sec:impl}
\begin{table}
	\centering
	\caption{Smartpick Properties}
	\label{tab:conf-em}
	\scalebox{0.7} {\begin{tabular}{lc}
			\hline
			Key&Default Value\\
			\hline
			smartpick.cloud.\textit{compute.provider} & AWS \\
			smartpick.cloud.\textit{compute.instanceFamily} & t3 \\
			smartpick.cloud.\textit{compute.relay} & True \\
			smartpick.cloud.\textit{compute.knob} & 0 \\
			smartpick.train.\textit{max.batch} & 100\\
			smartpick.train.\textit{pref.sameInstance} & False\\
			smartpick.train.\textit{min.ram.gb} & 4\\
			smartpick.train.\textit{errorDifference.trigger} & 50\\
			\hline
	\end{tabular}}
\end{table}

\label{sec:impl}
Smartpick is implemented on top of Spark 2.2.1 \cite{sparkRef}. 
Table \ref{tab:conf-em} shows Smartpick's properties that applications can easily set. 
Spark applications can easily utilize Smartpick by setting these properties without any modification. 
We will explain each property from the following explanation in detail.
Most components in Smartpick are implemented in Python 3.0 \cite{python} 
if not otherwise specified.  

\noindent\textbf{Workload prediction module:}
We designed and implemented the workload prediction module as a separate process (server) using Thrift RPC \cite{thrift}. 
Thus, other SEDA systems can get benefits from Smartpick, i.e., workload prediction and the cost-performance tradeoff feature.
We will show how two recent SEDA systems, 
i.e., Cocoa and Smartpick, utilize Smartpick as an external prediction system in Section \ref{subsubsec:stateArtComp}.

\noindent\textbf{Training prediction model:} To kick-start Smartpick, the first model training is invoked through a 
CLI (Command Line Interface) script, tailor-made to initialize and create models from scratch. 
When a prediction model needs to be trained either initially or in handling dynamics, 
we devise a heuristic to vary each training sample in the range of $\pm$ 5\% and create a reasonable dataset comprising around 10x samples (x being the original size). 
This task ensures that Smartpick can function quickly and effectively with as small as 100 representational workloads. 
Finally, the data burst is preceded and succeeded by random shuffling so that eventually, when the entire dataset is split into training and test sets, 
an unbiased selection is performed \cite{reshuffle}.

\noindent\textbf{Optimal cloud configurations}:
To determine the optimal cloud configuration with the prediction, \textit{compute.knob} can be set. 
If the best performance is preferred regardless of cost, it can be set to 0. Or it can be set any greater number than 0 to explore the cost-performance tradeoff space discussed in Section \ref{subsec:tradeoff}. 
Applications can set \textit{compute.instanceFamily} property to increase memory locality for further performance improvement, as discussed in Section \ref{sec:related}. 

\noindent\textbf{Query similarity check:} To parse the alien queries, the similarity checker (SC) uses the 
sql-metadata library \cite{sql-metadata}, which helps extract meaningful information such as the number of tables, columns and subqueries inferred in the request. 
Next, a 4-dimensional list is computed having all of the features (along with the number of map tasks), followed by the determination of spatial cosine similarity with 
respect to the known queries that helps filter out the best match. Thus, the closest query identifier is returned to the WP module, 
which then uses it to deduce the request's resource-needs.

\noindent\textbf{Prediction model updates:} Background re-training is necessary when the model is out of course and the predictions 
deviate from actual values beyond a pre-defined threshold, i.e., \textit{errorDifference.trigger}. 
An independent monitor thread in the MFE evaluates this condition and if required, creates a new model with \textit{warm\_start}, which is built as a pickle object for up-to-date reference. 
On completion, the monitor replaces this model in the referred directory, and all new workload predictions point to this object.
Smartpick allows users to select where the new model will be trained based on user's preferences, 
i.e., \textit{pref.sameInstance} and \textit{min.ram.gb}. If the same instance re-training is configured (\textit{pref.sameInstance}) and minimum memory (\textit{min.ram.gb}) is available, Smartpick spawns a new sub-process for re-training. Otherwise, a new instance is started and used for this purpose.
Smartpick also supports batch-based re-training (batch size given by the key \textit{max.batch}) that works independently to keep the model incrementally up-to-date.

\noindent\textbf{Metrics collection and history server:} 
To capture the metrics outlined in Table \ref{tab:fea-wp}, 
Spark's implementation of listener classes (along with the dependent modules) are modified 
and monitoring data is stored in JSON format. 
Once this model is in place, any subsequent request for data processing triggers asynchronous 
system-level events that have no (little) overhead to the ongoing job. 
The history server provides internal DNS (Domain Name System) as APIs for other components, e.g., MFE, to request and process the targeted metrics.

\noindent\textbf{Managing compute instances:} Resource manager (RM) is implemented on JDK 8 \cite{java8} using SDK libraries of AWS \cite{awsSDK} and Google Cloud \cite{gcpSDK}. 
Applications can point to the primary cloud provider by setting a Smartpick property - \textit{compute.provider}.
RM communicates with the respective cloud interface and launches the determined numbers of VMs and SLs.
Once these instances are up and running, it tracks their charging statuses for statistics on cost monitoring to be used later for performance/cost evaluation.

\noindent\textbf{Relay-instances mechanism:} 
To make the relay-instance mechanism active, the property \textit{compute.relay}
can be set to ``True''. SLs are terminated when relayed VM instances are ready to execute tasks.
To this end, RM will use mapping between REQUEST ID (for SL) and INSTANCE ID (for VM) after 
sending requests to cloud providers. 
When a VM instance is ready to be used and connects to RM with its INSTANCE ID, 
RM will find the corresponding target SL (REQUEST ID) using INSTANCE ID and let the task scheduler stop assigning tasks to it. 
After checking that no task is running on the SL, RM sends a termination message to it. 

\noindent\textbf{Cost estimation:} To estimate the cost for queries, we modified Spark workers to
send instance information such as ID, cloud provider, region,
type, storage type, and storage size to the RM when they
connect to it. While most information is static, thus hard-coded 
in the images, IDs are generated dynamically when
Smartpick sends requests to cloud providers, e.g., REQUEST ID
for SL and INSTANCE ID for VM. To identify each
worker, a boot script for VM and a function code for SL
acquire these IDs and set them as an environment variable. Using
these IDs, Smartpick tracks instances’ execution time and calculates 
overall compute resource cost for queries.
Since VM instances are charged only when they are in the
“Running” state, Smartpick uses a dedicated thread that checks
their statuses. In our implementation, each VM instance uses 8
GB (SSD) storage which is charged per second. While
SL does not charge for its volatile storage (2048 MB),
the external storage cost, e.g., AWS t3.xlarge or GCP e2-standard-4 for Redis, is added to the total cost if at least one SL instance
is running for a query. Note, data transfer within a DC is free
of charge in most cloud providers. 
\section{Evaluation} \label{sec:eval}
In this section, we present a detailed discussion of our evaluation to show the efficacy of Smartpick. 
\subsection{Experimental Setup}\label{subsec:setup}
\noindent\textbf{Compute resource setting}: We deployed Smartpick prototype implementation on 
live test-beds of AWS \cite{awsRef} (US East region) and GCP (US East region) \cite{gcpRef}.
On AWS, we use \textit{t3.xlarge} instance (4 vCPUs and 16 GB RAM) for the Spark master, Spark driver, and the external Redis server. For workers that are dynamically deployed at run-time, 
we use \textit{t3.small} instances (2 vCPUs and 2 GB RAM)
for VM and Lambda \cite{lambdaRef} 2 GB RAM for SL.
Note that each Lambda instance provides 2 vCPUs. 
That is, each VM and SL instance offer the same amount of
CPU cores and memory in our evaluation. On GCP, 
we use a similar compute resource setting to AWS, 
i.e., \textit{e2-standard-4} (4 vCPUs and 16 GB RAM) for the master, the driver, and the Redis server, and \textit{e2-small} (2 vCPUs and 2 GB RAM) and Function \cite{cloudFunctionsRef} with 2 GB RAM for workers. All experimental results are an average of 10 runs, plotted with 90\% confidence intervals.
For cost, we use cost information on AWS and GCP web pages for VMs and SLs. We consider storage cost, e.g., local disk storage of VM and external storage (Redis) instance for communication among SLs 
as explained in Section \ref{sec:impl}. 
We also consider burstable costs of \$0.05 per vCPU-hour as we use the \textit{t3} instance family. Note that burstable costs of GCP e2-small is free of charge, but users cannot control it.

\noindent\textbf{Applications}: For workloads to evaluate Smartpick, we use three popular benchmarks, TPC-DS \cite{tpcds}, TPC-H \cite{tpch}, and Word Count (WC) \cite{wordCountInHive}. 
TPC-DS suite comprises compute and I/O intensive workloads with a high number of dependent map and shuffle stages (6 $\sim$ 16). TPC-H benchmark has SQL-like query benchmarking (moderated compute and I/O) with a lesser sequence of stages (2 $\sim$ 6). Lastly, we use Word Count as a simple query with I/O requirement. 
For input data, we generate 100 GB of data in both AWS S3 and Google storage for each benchmark. 
While we observed similar patterns of results from these benchmarks, we mainly show the results from TPC-DS queries due to space constraints.
We use WC and TPC-H benchmarks as new queries to evaluate Smartpick's performance on workload dynamics. 
In addition, we generate separate 500 GB data for benchmarks to see how Smartpick reacts with changes to data size.

\noindent\textbf{Baselines}: 
We compare Smartpick's hybrid approach with two extreme approaches, i.e., SL-only and VM-only. 
To mimic VM-only and SL-only approaches, we tweak Smartpick's workload prediction module to choose either SL-only or VM-only for comparison purposes.
For the baselines, we compare the Smartpick against two state-of-the-art serverless-enabled data analytics systems, Cocoa \cite{cocoa} and SplitServe \cite{splitserve}. 
Note that we obtained the source code of Cocoa and SplitServe and integrated them into Smartpick's implementation on Spark for seamless comparisons.

\noindent\textbf{Building Prediction Models}: 
To train the prediction models, we run 20 randomly selected configurations of VMs and SLs 
for each of the 5 TPC-DS queries i.e., 11, 49, 68, 74, and 82, as representational workloads, short-, mid-, and long-running queries. 
We generate 1000 data samples, i.e., different SLs + VMs configurations, by the heuristic approach discussed in Section \ref{sec:impl}.
We use 800 samples to build prediction models and 200 samples to evaluate the accuracies of the models (Section \ref{subsec:workload}).
We build two prediction models, \textit{Smartpick} without relay-instances and \textit{Smartpick-r} with the relay-instances for comparison purpose.

\begin{table}[!t]
	\centering
	\caption{Performance comparison between GCP and AWS}
	\label{tab:compAWSGCP}
	\scalebox{0.65} {\begin{tabular}{|p{1.3cm}|p{2.3cm}|p{1.4cm}|p{1.2cm}|p{1.6cm}|p{1.4cm}|p{1.4cm}|}
			\hline
			\textbf{Provider}&\textbf{Cloud Storage (MiB/s)}&\textbf{VM I/O (writes/s)}&\textbf{VM I/O (reads/s)}&\textbf{Memory (1k-ops/s)}&\textbf{VM CPU (events/s)} & \textbf{SL CPU (events/s)} \\
			\hline
			AWS& 117.53& 771.06& 1156.59& 4675.66& 1109.07 & 811.13\\
			\hline
			GCP& 51.64& 764.14& 1146.21& 4182.49& 906.67 & 714.87\\
			\hline
	\end{tabular}}
\end{table}

\noindent\textbf{Performance Comparison between AWS and GCP:} 
To clearly understand the experimental results, we first describe the performance difference between AWS and GCP. 
Table \ref{tab:compAWSGCP} shows benchmark results between AWS and GCP; S3 and Storage for cloud storage, t3.small and e2-small for VM,
and Lambda and Function for SL.
Both of these VM and SL compute resources have 2 GB memory with dual vCPUs.
In order to collect the bandwidth information for Cloud Storage accesses, we upload a 1.5 GB text file onto AWS S3 and GCP Storage and then capture the time taken for download through a Python \cite{python} script. For the remaining measures, we use the Sysbench \cite{sysbench} with identical parameters on both the cloud providers. 
The table shows that AWS S3 provides better data transfer rate (bandwidth), which can affect overall query performance as input data is read from these cloud storage.
For CPU performance on VM, i.e., I/O, Memory, and VM CPU, AWS offers better performance than GCP. We observe that there is no significant difference in the boot-up time of VM as both require 31 $\sim$ 32 seconds. Similarly, for CPU comparisons on SLs, AWS offers better performance than GCP.
Additionally, SL workers on GCP \cite{cloudFunctionsRef} do not have ephemeral storage for source files other than the configured RAM \cite{gcpFunctionsVolInfo}, which further reduces the available memory for computation. 
In summary, the query execution times in GCP are comparably higher than that in AWS, which offers better performance for cloud resources we used in our evaluation. 
\begin{figure}[t!]
	\subfigure[Smartpick]{\includegraphics[width=1.6in]{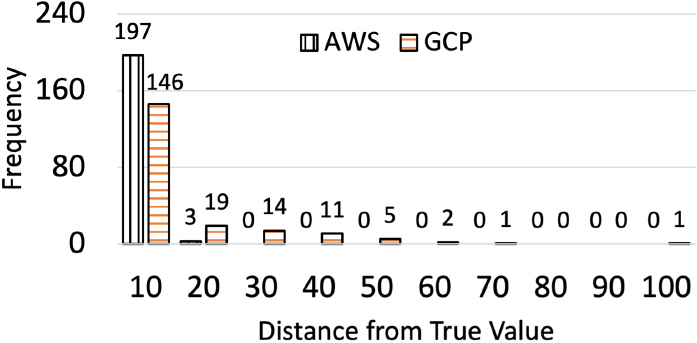}}
	\hfill
	\subfigure[Smartpick-r]{\includegraphics[width=1.6in]{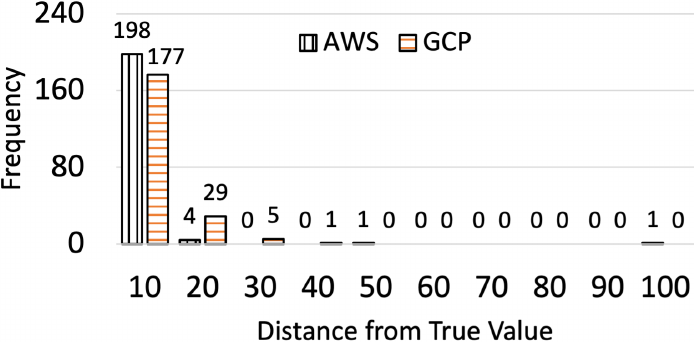}}
	\caption{Accuracy on test dataset}
	\label{fig:smartpickAcc-aws}
	\Description{The 2 sub-figures (a) and (b) show the validation of trained models, Smartpick and Smartpick-r, on the test dataset of AWS and GCP respectively. Both these figures suggest that the trained models accurately predict the final values in the test dataset.}
\end{figure}
\begin{figure*}[t!]
	\subfigure[Performance Comparison]{\label{subfig:awsperformance}\includegraphics[width=1.6in]{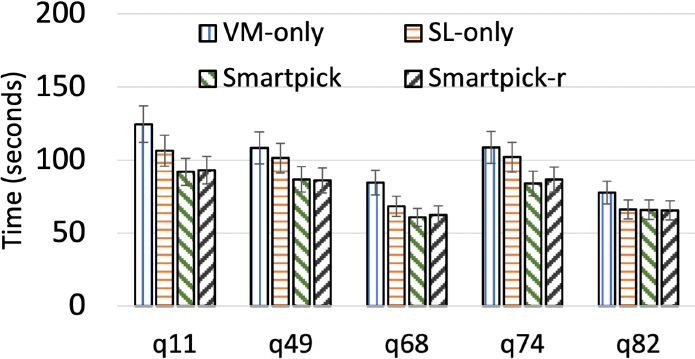}}
	\hfill
	\subfigure[Cost Comparison]{\label{subfig:awscost}\includegraphics[width=1.6in]{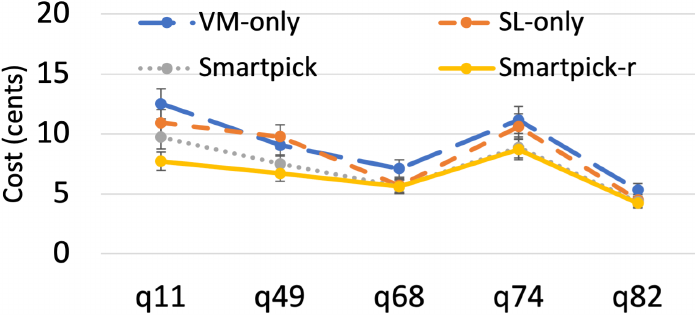}}
	\hfill
	\subfigure[Accuracy of Smartpick]{\label{subfig:awsacc}\includegraphics[width=1.6in]{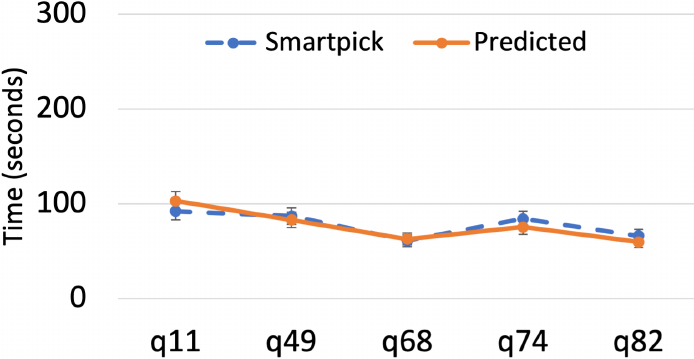}}
	\hfill
	\subfigure[Accuracy of Smartpick-r]{\label{subfig:awsaccr}\includegraphics[width=1.6in]{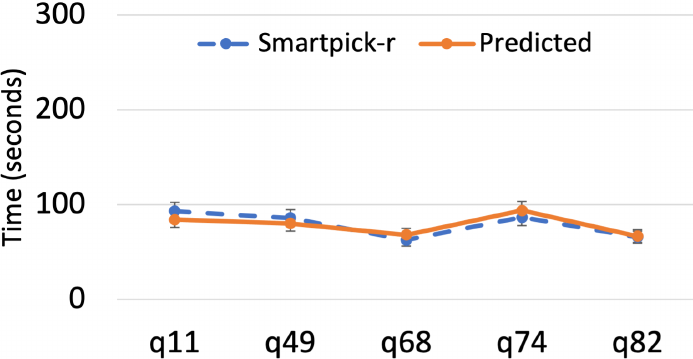}}
	\caption{Evaluation on AWS. (a), (b) - Lower is better. (c), (d) - Compactness is better.}
	\label{fig:smartpickNAWS}
	\Description{The figures (a) and (b) show the behavior of VM-only, SL-only, Smartpick and Smartpick-r with regards to two parameters, that is, performance and cost on AWS. It is observed that for all queries, Smartpick-r provides best performance at lowest cost. Also, figures (c) and (d) indicate high accuracy of Smartpick and Smartpick-r models respectively in predicting the actual query completion times.}
\end{figure*}
\begin{figure*}[t!]
	\subfigure[Performance Comparison]{\label{subfig:gcpperformance}\includegraphics[width=1.6in]{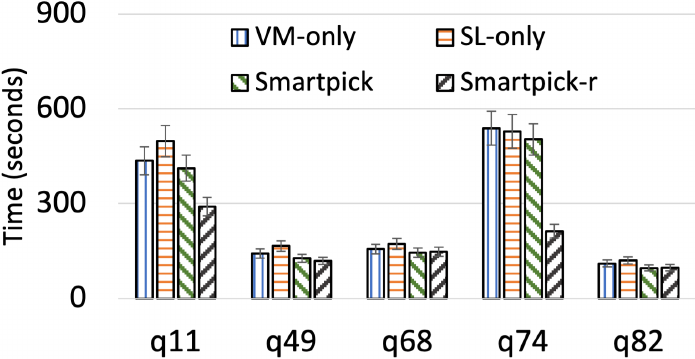}}
	\hfill
	\subfigure[Cost Comparison]{\label{subfig:gcpcost}\includegraphics[width=1.6in]{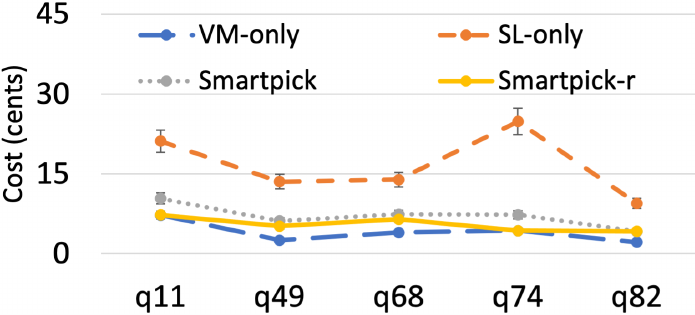}}
	\hfill
	\subfigure[Accuracy of Smartpick]{\label{subfig:gcpacc}\includegraphics[width=1.6in]{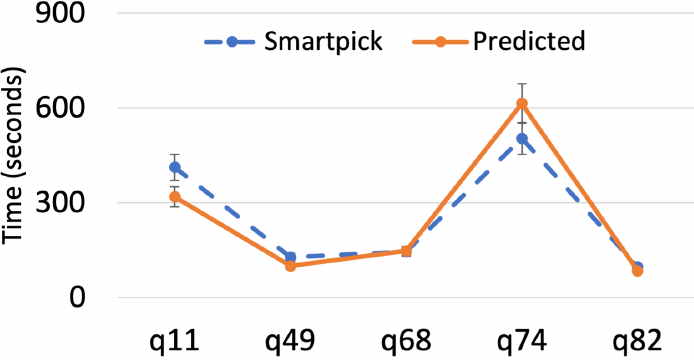}}
	\hfill
	\subfigure[Accuracy of Smartpick-r]{\label{subfig:gcpaccr}\includegraphics[width=1.6in]{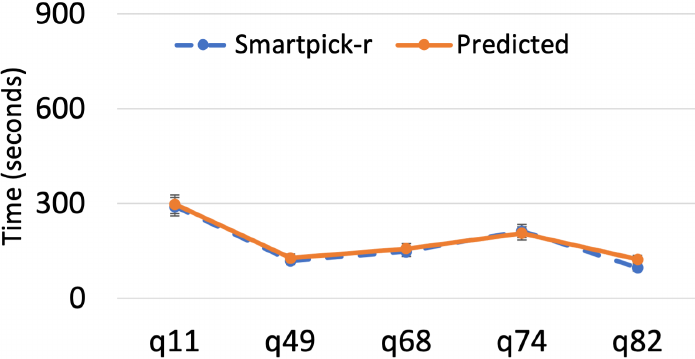}}
	\caption{Evaluation on GCP. (a), (b) - Lower is better. (c), (d) - Compactness is better.}
	\label{fig:smartpickNGCP}
	\Description{The figures (a) and (b) show the behavior of VM-only, SL-only, Smartpick and Smartpick-r with regards to two parameters, that is, performance and cost on GCP. It is observed that for all queries, Smartpick-r provides best performance at lowest cost. Also, figures (c) and (d) indicate high accuracy of Smartpick and Smartpick-r models respectively in predicting the actual query completion times.}
\end{figure*}
\subsection{Workload Prediction}\label{subsec:workload}
In this experiment, we show how accurately Smartpick and
Smartpick-r models predict given queries' workloads with the 
initial prediction models explained in Section \ref{subsec:setup}. We capture different key statistics from the model training phase. First, we see a reasonable Root Mean Squared Error (RMSE) for both the models, i.e., 
Smartpick and Smartpick-r. 
On AWS, we get RMSE scores of $6.2$ and $8.2$ respectively, 
where as on GCP, we get the same as $12.8$ and $7.59$ respectively. Based on the extensive statistical analysis, 
we take 2 times the standard error as an accurate enough prediction, since it considers both the directions of error (positive and negative) \cite{stderr}. 
Thus, we plot graphs for each of the above cases by 
considering the distance from truth values on the test dataset.

Figure \ref{fig:smartpickAcc-aws} shows the frequency of test samples (200/1000 in our experiments with an 80:20 hold-out split for training and testing respectively) at varying distances 
from the truth values in seconds. It is observed that for Smartpick on AWS, $98.5$\% of the predicted samples lie within $10$ seconds difference of the actual query execution times, 
which shows that the model yields accurate predictions \cite{stderr}. Likewise, Smartpick-r provides a prediction accuracy of $97.05$\% on AWS. 
Smartpick and Smartpick-r on GCP give prediction accuracies of $73.4$\% and $83.49$\%, respectively, which is due to higher query execution time on GCP that
incurs more variance. We assume that these results are reliable enough for prediction systems \cite{predAcc1, predAcc2, predAcc3, predAcc4}. 
Besides, the prediction model will become more accurate as Smartpick considers workload dynamics (Section \ref{wordCount}). 
\subsection{Performance and Cost Comparisons}\label{subsec:eval_compare}
\subsubsection{Comparisons with other approaches}
\label{subsub:compare_other approach}
In this experiment, we compare the performance of Smartpick and Smartpick-r to two baselines, i.e., VM-only and SL-only approaches. Note that the cost-performance knob ($\epsilon$) in this experiment is set to 0, i.e., the best performance. Figure \ref{fig:smartpickNAWS} shows the
results on AWS. Figure \ref{subfig:awsperformance}
and Figure \ref{subfig:awscost}
show query completion times and 
cost, respectively for five TPC-DS queries (11, 49,
68, 74, and 82) with 4 different approaches, i.e., VM-only, SL-only, Smartpick, and Smartpick-r. The results clearly show that both Smartpick models
achieve better performance to that of VM-only and SL-only approaches
with reduced cost. While we can see similar performance from Smartpick and Smartpick-r, Smartpick-r incurs less cost as expensive SLs
are terminated when corresponding VMs are ready, which shows the benefits of the \textit{relay-instances} mechanism. 
Figure \ref{subfig:awsacc} and Figure \ref{subfig:awsaccr} 
show predicted and actual query completion times using Smartpick and Smartpick-r respectively. 
These figures show that Smartpick can predict given queries' execution times accurately.
\begin{figure*}[t!]
	\subfigure[Performance on AWS]{\label{subfig:sysPerfAWS}\includegraphics[width=1.6in]{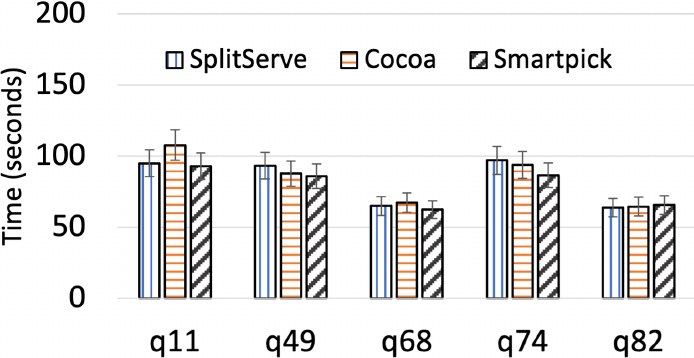}}
	\hfill
	\subfigure[Cost on AWS]{\label{subfig:sysCostAWS}\includegraphics[width=1.6in]{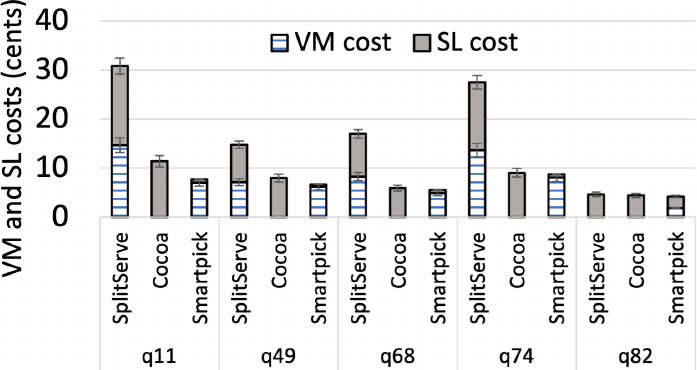}}
	\hfill
	\subfigure[Performance on GCP]{\label{subfig:sysPerfGCP}\includegraphics[width=1.6in]{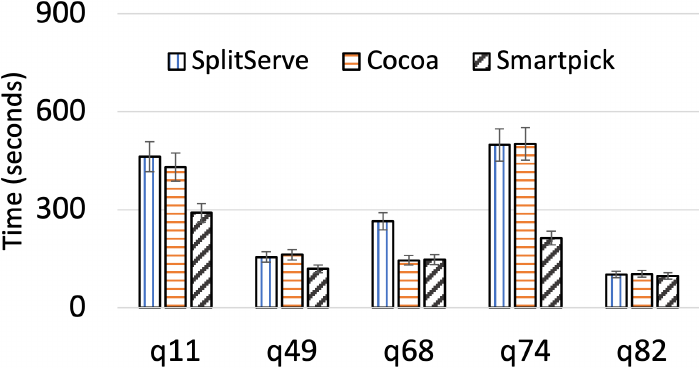}}
	\hfill
	\subfigure[Cost on GCP]{\label{subfig:sysCostGCP}\includegraphics[width=1.6in]{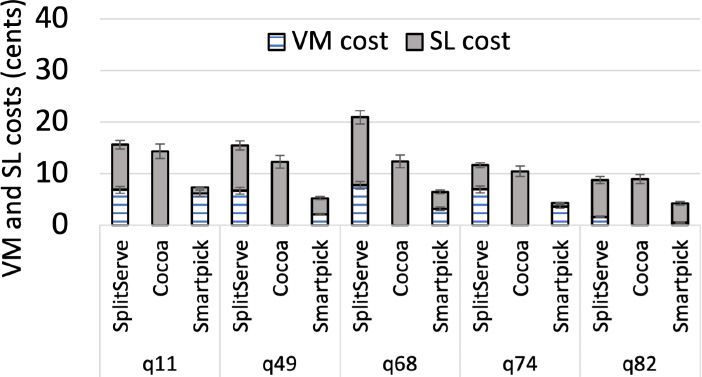}}
	\caption{Performance and cost comparisons with state-of-the-art (Cocoa and SplitServe). Lower is better.}
	\label{fig:smartpickSystems}
	\Description{Figures (a) and (b) compare the performance and cost of Smartpick-r to that of SplitServe and Cocoa on AWS. Figures (c) and (d) compare the same on GCP. It is observed that Smatypick-r provides the best/similar performance at lowest cost.}
\end{figure*}
Figure \ref{fig:smartpickNGCP} shows the similar patterns of results on GCP with more variance than AWS due to the different performance characteristics as explained in Section \ref{subsec:setup}.
For query 49 on GCP, we see a slightly better performance/cost compared to other queries, 
which is due to the persistent behavior of workload and significantly lesser variance. 
The VM-only cost on GCP is lower than other approaches as the burstable feature is free of charge on GCP. 
Overall, Smartpick-r shows better/similar performance with reduced cost compared to other approaches. In the rest of experiments, we use Smartpick to refer to Smartpick-r, unless otherwise mentioned.

\subsubsection{Comparisons with State-of-the-art Systems}\label{subsubsec:stateArtComp}
\label{subsub:compare_state_of_art}
In this section, we compare Smartpick with state-of-the-art systems, i.e., Cocoa \cite{cocoa} and SplitServe \cite{splitserve}. 
Since they rely on external workload prediction (WP) systems, we tweak our WP module to choose VM instead of SL + VM, 
and plug-in the module into Cocoa and SplitServe respectively as we discussed in Section \ref{sec:impl}.
Figure \ref{fig:smartpickSystems} shows the evaluation on AWS and GCP. 
We observe that SplitServe tends to give comparable query completion times, but at a high cost (VMs and SLs combined)
due to the underlying design of segueing, i.e., the same number of SL and VM, and static timeout threshold for SL, as we discussed in Section \ref{subsec:relay}.
Similarly, Cocoa gives comparable query completion times, but 
we see inflated costs for Cocoa as well.
This is because Cocoa tends to always favor SLs because of 
its dependency on other simply assumed static values, such as the execution time for each map/shuffle task, which significantly affects their decisions.
Thus, Smartpick can offer better query completion times with much reduced cost than other systems.


\begin{figure}[t!]
	\subfigure[Smartpick]{\label{subfig:trSR}\includegraphics[width=1.6in]{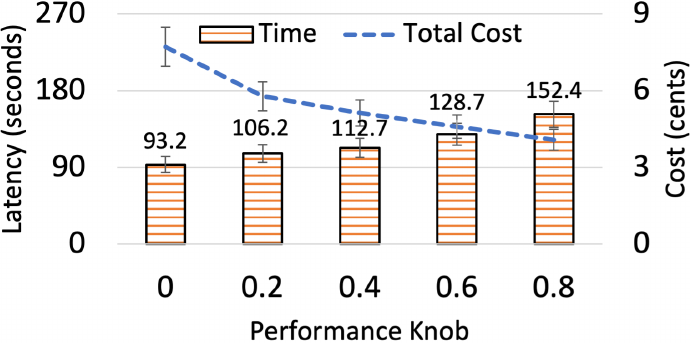}}
	\hfill
	\subfigure[SplitServe]{\label{subfig:trSS}\includegraphics[width=1.6in]{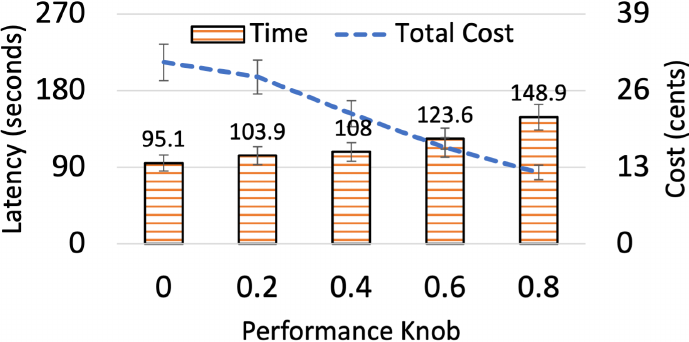}}
	\caption{Cost-performance tradeoff on AWS}
	\label{fig:troff-aws}
	\Description{Figure (a) shows the behavior of Smartpick-r with different values of the performance knob. As the knob is increased, that is higher tolerance to latency, cost is reduced significantly. Figure (b) shows the behavior of SplitServe with increasing values of the knob. Although a similar trend of performance and cost is observed, this figure helps illustrate the power of Smartpick for exploiting tradeoff spaces in other state-of-the-art systems.}
\end{figure}


\subsection{Exploiting cost-performance tradeoff}\label{subsec:tradeoffspaceEval}
For applications that have a limited budget, achieving the target performance
goal with the minimum cost is an important task, as discussed in Section \ref{subsec:tradeoff}. 
In this experiment, we show how such applications
achieve their cost-performance goals using Smartpick's property 
\textit{compute.knob}. Additionally, systems, e.g., SplitServe \cite{splitserve} that did not account for cost-performance tradeoff, can also benefit from Smartpick's design. Figure \ref{fig:troff-aws} shows the behavior of Smartpick and SplitServe (for query 11) with different values of the newly introduced performance knob. As applications increase the value of this knob from 0.2 - 0.8, the cost reduces significantly by trading off the query latency, as discussed in Section \ref{subsec:tradeoff}. 
Figure \ref{subfig:trSS} also shows that other systems, e.g., SplitServe, can benefit from Smartpick's feature by exploring the cost-performance tradeoff space.
Note that we could see a similar pattern of results from other queries on AWS and GCP, but omitted to cite these results here due to space constraints.

\subsection{Handling Dynamics}\label{handlingDynamicsParent}
\subsubsection{New Queries from TPC-DS workload}\label{subsubsec:simCheckEvalChild} 
\begin{figure}[t!]
	\subfigure[New TPC-DS queries on AWS]{\label{subfig:newTPCAWS}\includegraphics[width=1.6in]{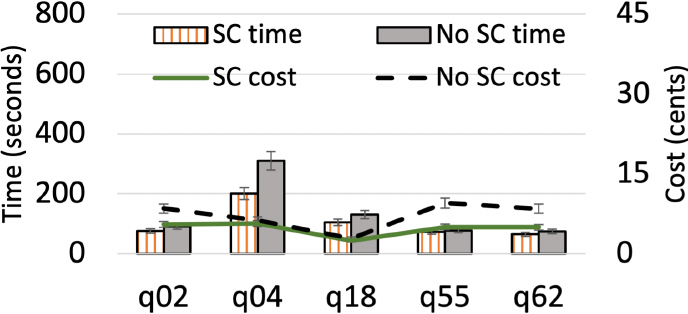}}
	\hfill
	\subfigure[New TPC-DS queries on GCP]{\label{subfig:newTPCGCP}\includegraphics[width=1.6in]{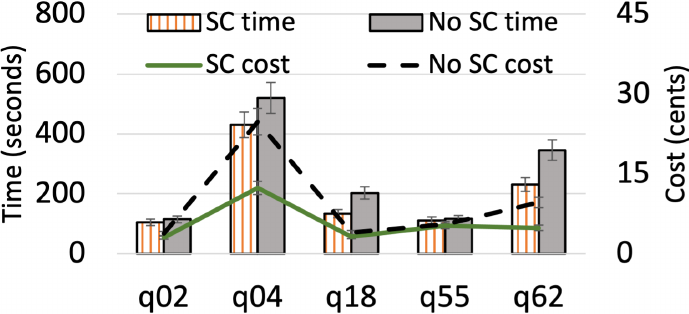}}
	\caption{Behavior with new TPC-DS queries}
	\label{fig:newTPCDS}
	\Description{Figures (a) and (b) show the efficacy of Smartpick in dealing with similar but untrained TPC-DS queries on AWS and GCP respectively. It is evident that Smartpick with SC module outperforms the vanilla prototype (without SC) and generates best query performance at lowest cost.}
\end{figure}

To see how Smartpick handles other 
queries of TPC-DS, we use the queries 2, 4, 18, 55, and 62,
as unknown queries to Smartpick that have similar workloads 
with the queries used for building prediction models. Figure \ref{fig:newTPCDS} shows the benefit from Similarity Checker (SC), which helps achieve the best query latency ($\epsilon=0$) at a reduced cost for all new queries. This highlights the significance of SC module for similar workloads, which was discussed in detail in Section \ref{sec:design}. 

\begin{figure}[t!]
	\subfigure[Word Count on AWS]{\label{subfig:wrdAWS}\includegraphics[width=1.6in]{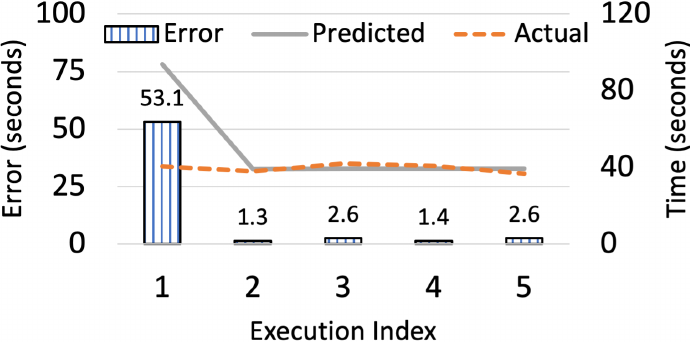}}
	\hfill
	\subfigure[Word Count on GCP]{\label{subfig:wrdGCP}\includegraphics[width=1.6in]{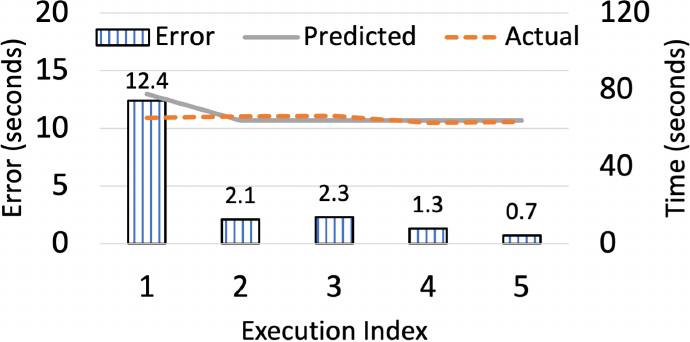}}
	\caption{Word Count problem on Smartpick}
	\label{fig:wrdCnt}
	\Description{Figures (a) and (b) show the efficacy of Smatpick-r in handling new workloads such as word count problem on AWS and GCP respectively. With background retraining, the model is able to converge quickly, and the error between predicted- and actual- query-completion time reduces after the first execution run. \textbf{KO: it would be better to compare the execution time and predicted time. Explain some number (cost and performance) in text}}
\end{figure}

\begin{figure}[t!]
	\subfigure[TPC-H on AWS]{\label{subfig:tpcAWS}\includegraphics[width=1.6in]{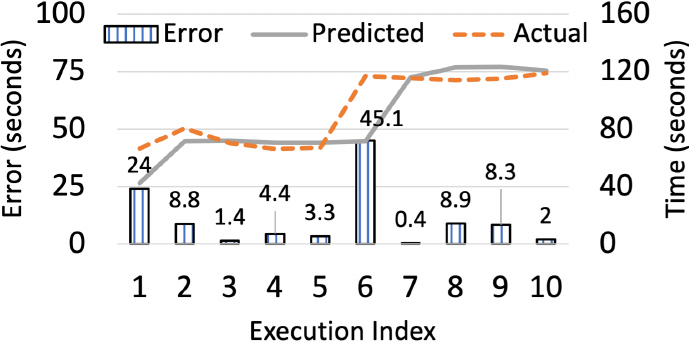}}
	\hfill
	\subfigure[TPC-H on GCP]{\label{subfig:tpcGCP}\includegraphics[width=1.6in]{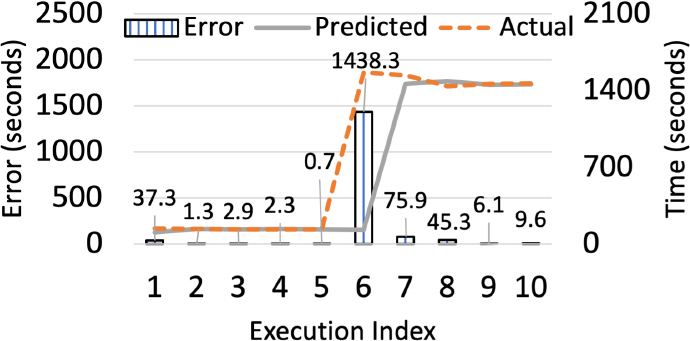}}
	\caption{TPC-H on Smartpick with change in data size}
	\label{fig:tpch}
	\Description{Figures (a) and (b) show the efficacy of Smartpick-r in quickly adapting to increased data sizes on AWS and GCP respectively. Although there is spike in error for sudden increase in data size, the proposed prototype quickly learns the nature of changes through background re-training. This reduces the error for subsequent runs and the model converges faster.}
\end{figure}

\subsubsection{Handling new workloads and increase in size}\label{wordCount}
One of the key aspects of Smartpick is to handle new queries 
by retraining models with the characteristics of new workload. 
In this section, we use Word Count (WC) as a new workload to
Smartpick. Based on the early trials, we observe that same instance re-training leads to an overhead on the ongoing job (which is expected), and therefore, advocate the use of different instance re-training (unless required otherwise).
To trigger the model retraining, we set \textit{errorDifference.trigger} to 10. That is, if the difference between actual query execution time 
and predicted time is more than 10 seconds, then model retraining is triggered. When the new query is submitted at first, 
Similarity Checker is invoked for each unknown query and the job proceeds to termination based on the closest match 
as discussed in Section \ref{subsec:designDynamics}. Upon job termination, an independent monitor thread triggers background re-training if the difference in predicted and actual values is higher than the configured threshold (\textit{errorDifference.trigger}). Figure \ref{fig:wrdCnt} shows that Smartpick's prediction model quickly converges to new values by efficient (data-burst based) re-training, as discussed in Section \ref{subsec:designDynamics}.

Another important aspect of handling dynamics is the change in workload size. We follow the same set-up as above, but instead use TPC-H query 3 workload as an alien query. 
In addition, after 5 executions, we change the database to point to a larger size of 500 GB and clean the event logs for existing query. 
While such significant changes may be rare in real environment, the dataset size grows eventually and consistently with increasing use of the application.
Figure \ref{fig:tpch} shows the results observed for query 3. Clearly, when the data size shoots up, 
Smartpick can capture this change and quickly converges to the actual execution times. 
This support of handling dynamics asynchronously and quickly makes Smartpick a robust application with enhanced reliability even in the presence of workload dynamics. Note that the larger spike in the case of GCP is because of the slowness of cloud resources (as discussed in Section \ref{subsec:setup}), which is further aggravated by the large input data size of 500 GB.
\section{Related Work} \label{sec:related}

\noindent\textbf{Exploiting SL and VM together}:
LIBRA \cite{libra}, aims to reduce the cost of hybrid deployments by utilizing cost indifference point, though actual costs can vary depending on the granularity of estimated completion time, where Smartpick comes into play. Cocoa \cite{cocoa} depends on static parameters and does not support relaying of workloads from SLs to VMs, which results in inflated cost. 
While SplitServe \cite{splitserve} incorporates segueing from SLs to VMs, it results in cost inflation due to its design. 
It also demands the end-user to employ a cost manager for determining the additional SL resources, which is burdensome work.
SplitServe \cite{splitserve}, MArk \cite{perf-cost-ratio}, FEAT \cite{robustScaling} and Spock \cite{slo-spock} aim at reactively launching the SL instances whenever free cores are unavailable. Conversely, Smartpick's resource determination scheme optimizes the choice of VMs and SLs together while meeting cost-performance goals.


\noindent\textbf{Workload prediction for compute resource configurations:} 
Numerous prior works \cite{ernest, vanir, selecta, paris, optimus, cherrypick, optimuscloud, crystalLp, rpss, juggler}
have proposed methodical workload prediction schemes that help determine resource configurations for VM-based workloads. Adding SLs to the supported compute instance types leads to a huge search space for optimality and thus, renders these techniques time-consuming and ineffective. Interestingly, PerfOrator \cite{PerfOrator} uses hardware-level statistics to build performance model of big data queries, whereas Smartpick requires no advance knowledge of hardware settings and even supports the hybrid model of SLs and VMs.

\noindent\textbf{Handling dynamics:} CherryPick \cite{cherrypick} relies solely on the BO model to incorporate cloud uncertainties into the decision-making. This works fine for VM instance families but is not well suited to the hybrid approach for ad-hoc alien queries. Jockey \cite{jockey}, Morpheus \cite{Morpheus}, and ARIA \cite{aria} dynamically tune resource allocations (based on historical data) to ensure time-critical jobs with stringent SLOs are provided with required compute resources. They are, however silent on types of compute resources and do not consider the cold boot-up time of VMs. Conversely, Optimus \cite{optimus} does not depend on the historical information and imposes a checkpoint-inspired technique to handle changes in parameter servers, which can lead to a huge overhead due to multiple restarts. Quasar \cite{Quasar} updates its (VM) resource allocation approach based on active monitoring and sensitivity of the application's performance. Smartpick, instead, can handle unknown requests by employing spatial cosine similarity and course-grained dynamics, as shown in Section \ref{subsec:designDynamics}.

\noindent\textbf{Enhancing memory locality}: 
Many serverless-enabled data analytics systems \cite{spock, splitserve, occupy, flint, pocket, locus, numpywren, spark_on_lambda}
have utilized external storage systems, such as Redis and AWS S3, to avoid SL's limitation, i.e., limited network. 
However, this may naturally cause performance degradation due to losing data (memory) locality.
Some recent works \cite{punching, boxer, short-lived, fmi} showed that
SL instances can communicate with each other directly using TCP hole punching and socket-related library replacement. 
We expect that using such techniques would improve performance for diverse queries, especially short-running queries.
We plan to apply these techniques in Smartpick for further performance improvement without additional cost.
To improve memory locality, we also consider using larger (expensive) VM instance types (and families).
We could observe that applications can improve performance with additional cost by using 
larger VM instance family, e.g., AWS c3, which opens another richer tradeoff space.
However, we omitted this result due to space constraints. 

\section{Conclusion} \label{sec:concl}
In this paper, we present Smartpick, a scalable data analytics system that determines optimal
compute resource configurations for given queries by predicting workloads with consideration of
hybrid compute resources, i.e., SL and VM.
Smartpick utilizes decision-tree based Random Forest to predict workloads
and Bayesian Optimizer to efficiently explore the large search space for determining optimal configurations.
Smartpick is mindful of cost-performance tradeoff space opened by exploiting SL and VM together, and incorporates workload dynamics.
Experimental results on AWS and GCP indicate high-precision resource determination for Smartpick with prediction
accuracies of 97.05\% and 83.49\% respectively. The results confirm that Smartpick enables 
applications to achieve their target cost-performance goals, handle workload dynamics automatically, and 
improve performance without additional cost compared to baselines.
The results also show that other data analytics systems can benefit from Smartpick.

\bibliographystyle{ACM-Reference-Format}
\bibliography{sample-base}

\appendix

%
%
%
%
%
%
%

\end{document}